\documentclass[conference]{IEEEtran}
\IEEEoverridecommandlockouts

\usepackage{cite}
\usepackage{amsmath,amssymb,amsfonts}
\usepackage{algorithmic}
\usepackage{graphicx}
\usepackage{textcomp}
\usepackage{xcolor}
\def\BibTeX{{\rm B\kern-.05em{\sc i\kern-.025em b}\kern-.08em
    T\kern-.1667em\lower.7ex\hbox{E}\kern-.125emX}}

\usepackage{comment}
\usepackage{booktabs}
\usepackage{color, colortbl}
\usepackage{pgfplots}
\definecolor{LightCyan}{rgb}{0.88,1,1}
    
\begin{document}

\title{Codec2Vec: Self-Supervised Speech Representation Learning Using Neural Speech Codecs}

\author{
\IEEEauthorblockN{1\textsuperscript{st} Wei-Cheng Tseng}
\IEEEauthorblockA{\textit{Department of Computer Science} \\
\textit{University of Texas at Austin}\\
Texas, USA \\
raytseng@utexas.edu}
\and
\IEEEauthorblockN{2\textsuperscript{nd} David Harwath}
\IEEEauthorblockA{\textit{Department of Computer Science} \\
\textit{University of Texas at Austin}\\
Texas, USA \\
harwath@utexas.edu}
}

\maketitle

\begin{abstract}
Recent advancements in neural audio codecs have not only enabled superior audio compression but also enhanced speech synthesis techniques. Researchers are now exploring their potential as universal acoustic feature extractors for a broader range of speech processing tasks. Building on this trend, we introduce Codec2Vec, the first speech representation learning framework that relies exclusively on discrete audio codec units. This approach offers several advantages, including improved data storage and transmission efficiency, faster training, and enhanced data privacy. We explore masked prediction with various training target derivation strategies to thoroughly understand the effectiveness of this framework. Evaluated on the SUPERB benchmark, Codec2Vec achieves competitive performance compared to continuous-input models while reducing storage requirements by up to 16.5× and training time by 2.3×, showcasing its scalability and efficiency.
\end{abstract}


\section{Introduction}
Over the past several years, the speech processing community has rapidly adopted self-supervised learning (SSL) followed by supervised fine-tuning as a general-purpose modeling approach for tasks ranging from automatic speech recognition and emotion recognition to speaker verification~\cite{ssl-review, superb, SUPERBJournal}.
Typically, the self-supervised phase involves pre-training a large-scale "foundation" model using substantial amounts of unlabeled audio data via pretext tasks such as masked prediction~\cite{Discrete-BERT, ling2020decoar, hubert, liu2024dinosr} or contrastive learning~\cite{cpc, wav2vec2}.
Due to the sheer volume of pre-training data, this phase is significantly more computationally expensive than fine-tuning~\cite{MelHubert, AcademicHuBERT, lugo2024sustainable}.
Moreover, this computational cost is further exacerbated as SSL models generally process audio input in the form of raw waveforms or high-dimensional continuous features like Mel spectrograms, requiring substantial processing and storage.

\begin{figure*}[!t]
    \centering
    \includegraphics[width=0.97\textwidth]{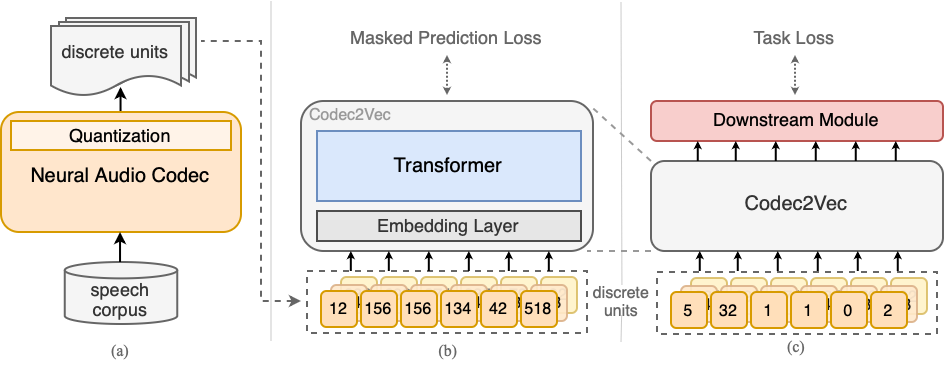}
    \caption{Overall pipeline of Codec2Vec. The pipeline comprises three stages: (a)Pre-computation of discrete codec units for the pre-training dataset using a neural audio codec; (b)~Codec2Vec SSL pretraining using masked prediction; (c)~after pretraining, fine-tuning with a lightweight downstream module using discretized dataset inputs.}
    \label{fig:pipeline}
\end{figure*}

Parallel to these SSL advancements, the speech and audio community has made notable progress in developing neural audio codecs with superior compression and fidelity over traditional algorithms~\cite{soundstream, Encodec, DAC, speechtokenizer}.
These codecs often employ vector quantization within the latent space of a neural autoencoder, compressing the input audio into discrete unit sequences.
Beyond compression, these discrete units have proven effective not only for high-quality speech synthesis~\cite{VALL-E, voicebox, voicecraft, speechx, zhan2024anygpt} but also show promise as alternative input features for broader speech processing tasks~\cite{codecASR}.
The appeal of codec units lies in their unique representational properties: 
compared to raw waveform or spectrograms, codec units are more compact, making them well-suited for scenarios where storage, transmission efficiency, or latency is a concern.
Moreover, recent work~\cite{codecSUPERB} also suggest that, despite quantization-induced lossy compression, codec units retain rich acoustic and linguistic information to support a wide range of speech processing applications.

Building on these observations, several works have investigated the use of codec units directly as model inputs for speech understanding tasks.
For instance, CodecASR~\cite{codecASR} and Puvvada et al.\cite{puvvada2024discrete} investigate replacing spectrograms with codec units for automatic speech recognition and speaker-related tasks, while VioLA\cite{viola} reformulates speech tasks as conditional language modeling over codec unit sequences.
Although these efforts report promising task-specific results, they rely on supervised training and do not offer a general-purpose representation learning framework.
Furthermore, benchmarks such as DASB~\cite{mousavi2024dasb} reveal a critical limitation: when used without additional learning, codec units alone underperform across a range of speech tasks, particularly those requiring contextual understanding.
This highlights a key challenge: while codec units encode important acoustic cues, they lack the contextual modeling needed to generalize across tasks.
Taken together, these results motivate a fundamental research question:
Can a general-purpose self-supervised speech representation model be trained directly on discrete audio codec units?

In this paper, we attempt to answer this question by introducing \textbf{Codec2Vec}, the first self-supervised speech representation learning framework that operates exclusively on discrete audio codec units.
Unlike conventional SSL approaches that extract features from continuous waveforms, Codec2Vec relies entirely on pre-computed discrete representations, marking a shift in input modality and enabling more scalable and modular learning.
Crucially, rather than proposing a new learning objective, our focus lies in demonstrating that standard SSL techniques, specifically masked prediction, remain effective when applied to this fully discrete input setting.
This exploration aims to deepen the understanding of the applicability and generalization ability of codec units across a broad spectrum of speech tasks~\cite{superb, SUPERBJournal, mousavi2024dasb}.

The Codec2Vec framework offers several practical advantages. 
First, using pre-computed compressed speech units significantly reduces data storage requirements, which is a critical factor for training on large-scale datasets, and facilitates efficient data transmission across networks, such as in distributed training.
In some cases, it also enables pre-training datasets to be entirely loaded into RAM, substantially mitigating I/O bottlenecks during training. 
Second, by eliminating the need for the convolutional waveform encoder during the SSL pre-training phase (after a one-time offline codec unit extraction), the computational cost and training time for the representation learning itself are largely reduced. 
Third, this framework can enhance data privacy, as reconstructing the original waveform from the discrete tokens is infeasible without access to the original codec model. 

Our contributions are summarized as follows:
\begin{itemize}
    \item We introduce \textbf{Codec2Vec}, the first exploration of self-supervised speech representation learning that relies exclusively on discrete units generated by neural audio codecs.
    \item We demonstrate that existing SSL mechanisms, specifically masked prediction, can be effectively applied to this discrete input modality, and we investigate the performance of various strategies for deriving training targets.
    \item We demonstrate that Codec2Vec achieves competitive performance across most tasks in the SUPERB benchmark compared to continuous-inputs models, highlighting the feasibility of using audio codec units for broad speech-processing applications.
    \item We show the substantial efficiency gains: Codec2Vec reduces data storage requirements by up to 16.5x and accelerates the SSL pre-training phase by up to 2.3x, highlighting its scalability for large-scale pre-training and resource-constrained environments.
\end{itemize}

\section{Related Works}
Over the past decade, self-supervised learning (SSL) has become a cornerstone of modern speech representation learning~\cite{ssl-review}.
Models such as HuBERT~\cite{hubert}, WavLM~\cite{wavlm}, and DinoSR~\cite{liu2024dinosr} have consistently demonstrated that pre-training on large volumes of unlabeled raw audio waveforms can yield powerful representations that generalize across a wide array of downstream tasks, significantly reducing the dependency on extensive annotated datasets.~\cite{superb,SUPERBJournal}.
However, the computational expense associated with pre-training these models, particularly due to the processing of high-dimensional continuous inputs (waveforms or spectrograms) by convolutional feature extractors, remains a significant challenge as data scales continue to grow. In response, one line of research has focused on improving efficiency by modifying these feature extractors or using lighter-weight continuous input features~\cite{MelHubert,SEW,InefficiencyAFE}.
Our work aligns with this research direction by proposing a more fundamental shift: eliminating the online acoustic feature extraction process altogether during SSL pre-training by leveraging pre-computed discrete audio codec units as the exclusive input.

Another closely related line of research involves SSL with discretized speech representations, often derived from the outputs of initial SSL models that were themselves trained on continuous audio.
For instance, DiscreteBERT~\cite{Discrete-BERT} and CoBERT~\cite{meng2022cobert} applied BERT-style masked language modeling to sequences of such discrete speech units.
Similarly, Chang et al.~\cite{chang2024exploring}  analyzed the performance of various discrete speech units derived from different SSL models across multiple speech understanding tasks. 
While these works demonstrate the utility of discretized inputs and provide valuable analyses of SSL-derived units, they typically rely on tokens generated by an SSL model that first processed continuous signals, thus inheriting the computational aspects of that initial stage.

Separate from these, another body of work has focused on leveraging the discrete units generated directly by neural audio codecs.
For example, EncodecMAE~\cite{pepino2023encodecmae}, utilize these codec units as prediction targets for a masked autoencoding task, though their primary encoder still processes continuous audio inputs to predict these discrete targets.
Other research explores the direct use of codec units as acoustic features for specific downstream applications. For example, Puvvada et al.~\cite{puvvada2024discrete} and CodecASR~\cite{codecASR} have built competitive speech recognition and speaker verification systems using discrete codec units as the primary input.
VioLA~\cite{viola} treats various speech processing tasks as conditional language modeling problems operating on sequences of codec units.
DASB~\cite{mousavi2024dasb} further facilitates the systematic evaluation of these diverse discrete audio units directly on a range of downstream tasks, contributing to a deeper understanding of their inherent capabilities and current limitations.
These studies demonstrate the rich information within neural audio codec units and their potentail for more general speech processing tasks.
However, their primary focus mostly lies in the direct application or task-specific modeling of these units, leaving open further exploration into how general-purpose representations can be self-supervised when relying exclusively on these discrete units as input.

Unlike these prior efforts, Codec2Vec advances a different paradigm: it aims to learn general-purpose speech representations by operating exclusively on discrete audio codec units as its input throughout the entire SSL pre-training phase.
This approach fully decouples the SSL model from continuous waveform processing during the representation learning stage, thereby maximizing the potential efficiency gains. 
By demonstrating that competitive representations can be learned directly from these codec-derived units, Codec2Vec paves the way for more scalable and resource-efficient speech foundation models.


\section{Methodologies}
\textbf{Codec2Vec} is a self-supervised representation learning framework that operates exclusively on discrete units generated by neural audio codecs.
Figure~\ref{fig:pipeline} presents an overview.
Following the standard "pretrain-then-finetune" paradigm in self-supervised learning (SSL)~\cite{ssl-review}, our approach present a paradigm shift by completely replacing conventional continuous inputs (like waveforms or spectrograms) with these pre-computed discrete units.
Specifically, we first compress all training audio using an off-the-shelf neural audio codec to obtain sequences of audio codec units. 
These units then serve as the sole input to our model during pre-training, where we employ a masked prediction objective to learn contextualized speech representations. 
After pretraining, a lightweight downstream module is appended to the encoder and fine-tuned on downstream tasks to evaluate the learned representations.
Below, we describe how audio codec units serve as model inputs, present the formulation of the masked prediction task, and describe several strategies for deriving training targets. 

\subsection{Using Discrete Inputs from Neural Audio Codecs}
To compress speech into audio codec units, we adopt DAC~\cite{DAC} due to its strong information-preserving capabilities~\cite{codecSUPERB}. We use the 16kHz variant, the most common sample rate in speech representation learning. DAC produces $N_{cb}=12$ codebook sequences at a rate of 50Hz. Formally, a speech signal $X$ is mapped to:
\begin{equation}
    \text{Codec}(X) = Q = \{Q_1, Q_2, \dots, Q_{N_{cb}}\}
\end{equation}
where each $Q_i = [q_{i_1}, q_{i_2}, \dots, q_{i_T}]$ is a sequence of discrete units, and $T$ is the total number of frames.
To feed these discrete units into our model, each code $q_{i_t}$ is transformed via an embedding table $E_i(\cdot)$
$Z_i = [\mathbf{z}_{i_1}, \mathbf{z}_{i_2}, \dots, \mathbf{z}_{i_T}]$.
We then aggregate these into a single sequence: 
\begin{equation}
Z = [\textbf{z}_1, \textbf{z}_2, \dots, \textbf{z}_T] = [\Sigma \textbf{z}_{i_1}, \Sigma \textbf{z}_{i_2}, \dots, \Sigma \textbf{z}_{i_T}].
\end{equation}
which serves as the input to the Transformer model.

In preliminary experiments, we found that initializing the embedding~\cite{codecASR} layer with the codec’s own codebook embeddings yields substantial performance gains. Additionally, applying quantizer dropout~\cite{soundstream} during training, where a subset of codebook sequences is randomly dropped with a specified probability, further improves robustness against noisy speech.

\subsection{Representation Learning with Masked Prediction}
To learn contextualized speech representations from the discrete codec unit sequences, Codec2Vec adapts the widely-used masked prediction objective, a technique proven effective in numerous self-supervised learning frameworks~\cite{ling2020decoar, hubert, liu2024dinosr}. Given the input vector sequence $Z$, we randomly select a subset of time indices $M \subset [T]$ and replace the corresponding frames with a learnable mask embedding, producing a corrupted sequence $\Tilde{Z}$.
The Transformer encoder, followed by a projection layer, then predicts the probability distribution over different targets, formulated as:
\begin{align}
    p_f(c|\Tilde{Z}, t) = \frac{\text{exp}(w_{c}^\top\mathbf{h}_t/\tau)}{\sum_{c'=1}^{C}\text{exp}(w_{c'}^\top\mathbf{h}_t/\tau)}
\label{equation:eq2}
\end{align}
where $c$ is the training target, $[\mathbf{h}_1, \mathbf{h}_2, \ldots, \mathbf{h}_T]$ symbolize the Transformer's output hidden representations $H$, $w$ is the projection layer, and $\tau$ is the temperature parameter. 

\begin{table*}[!t]
\centering
\caption{Evaluations on SUPERB benchmark. The tasks are categorized into four domains. ParaL. denotes paralinguistic tasks. Metrics for each task are: phone error rate (PR), word error rate (ASR), F1 and concept error rate (SF), equal error rate (ASV), diarization error rate (SD), and accuracy (KS, IC, and ER). *DeCoAR 2.0 introduces an additional vector quantization (VQ) layer between its encoder and reconstruction modules as part of its representation learning process. \textsuperscript{\textdagger}Pretrained using \textsc{HuBERT-iter0-layer9-kmeans500}, which is the same training target used for pretraining HuBERT~\cite{hubert}.}
\setlength{\tabcolsep}{4pt}
\begin{tabular}{lcccrrrrrrrrr}
\toprule
            & &  &  & \multicolumn{3}{c}{Content} & \multicolumn{2}{c}{Semantic} & \multicolumn{2}{c}{Speaker} & ParaL.  \\
            \cmidrule(lr){5-7} \cmidrule{8-9} \cmidrule(lr){10-11} \cmidrule(lr){12-12}
Method      & Input & Dataset size (GB)& Choice of Target & PR$\downarrow$      & ASR$\downarrow$     & KS $\uparrow$     & IC $\uparrow$     & SF(F1$\uparrow$/CER$\downarrow$)  & SD$\downarrow$ & SV$\downarrow$     & ER $\uparrow$    \\
\midrule
\multicolumn{2}{l}{\textit{reconstruction-based}} \\
DeCoAR 2.0~\cite{ling2020decoar}*  & spectrogram & 60.4& spectrogram & 14.9   & 13.0   & 94.5   &  90.8  & 83.3 / 34.7    & 6.6  & 7.2   & 62.5   \\
\rowcolor{LightCyan}
Codec2Vec  & discrete codes & 3.6  & discrete codes  & 19.6   & 13.9   & 93.6   & 78.3   & 83.6 / 34.2    & 5.6  & 6.8    & 59.1   \\
\midrule
\multicolumn{2}{l}{\textit{iterative clustering}} \\
HuBERT~\cite{hubert}      & waveform& 60.4 & k-means (HuBERT-iter0) & 5.4    & 6.4    & 96.3   & 98.3   & 88.5 / 25.2    & 5.9  & 5.1    & 64.9   \\
Codec2Vec\textsuperscript{\textdagger}  & discrete codes& 3.6 & k-means (HuBERT-iter0) & 5.2    & 6.9    & 96.7   & 98.0   & 88.6 / 24.5    & 5.4  & 5.1    & 64.9  \\
\rowcolor{LightCyan}
Codec2Vec & discrete codes& 3.6 & k-means (Codec2Vec) & 5.5 & 7.2    & 96.4   & 97.1   & 88.9 / 24.1    & 5.5  & 5.2   & 65.4   \\
\midrule
\multicolumn{2}{l}{\textit{online clustering}} \\
DinoSR~\cite{liu2024dinosr}  & waveform& 60.4 & EMA+VQ    & 3.2    & 4.7    & 96.7   & 98.0   & 88.8 / 23.6    & 5.4  & 5.5     & 65.9     \\
\rowcolor{LightCyan}
Codec2Vec & discrete codes& 3.6 & EMA+VQ & 4.2    & 6.2    & 96.7   & 98.4   & 88.7 / 24.4    & 5.5 & 6.0    & 64.9  \\
\bottomrule \bottomrule
\end{tabular}
\end{table*}

It is worth noting that the choice of training targets plays a crucial role for the learned representations. To fully understand the effectiveness of the Codec2Vec framework, we explore the following strategies for deriving the training targets:\\
\noindent \textbf{(a) Reconstruction-based.}  
The most straightforward approach is to task the model with reconstructing the original input audio codec units of the masked part.
This strategy is analogous to DiscreteBERT, which learns representation by performing masked language modeling on self-supervised speech units.
It also shares conceptual similarities with DeCoAR 2.0~\cite{ling2020decoar} and TERA~\cite{tera}, which reconstruct masked segments of local acoustic features, except our input is discretized.
In practice, we slightly adapt eq.~\ref{equation:eq2} with multiple projection layers to predict multiple codebook sequences during training.\\
\noindent \textbf{(b) Iterative Clustering.}  
Following HuBERT~\cite{hubert}, we refine our targets across successive training rounds by applying unsupervised clustering to the model’s intermediate latent representations. We start with a model pretrained using reconstruction-based targets and then repeatedly apply $k$-means on its latent representations to produce new frame-level $k$-means assignment for subsequent training.\\
\noindent \textbf{(c) Online Clustering.}  
Inspired by DinoSR~\cite{liu2024dinosr}, we also use an online clustering approach. Here, a “teacher” model dynamically learns a codebook from its own intermediate representations, generating cluster assignments for a “student” model to predict. The teacher is updated as an exponential moving average (EMA) of the student, providing continuously updated targets during training.

\section{Experiments}


Following prior works~\cite{hubert,liu2024dinosr}, we use 960 hours of speech data from the LibriSpeech corpus to pre-train our models. The model architecture is a \textsc{Base}-sized Transformer Encoder, consisting of 12 layers with an embedding dimension of $768$.
For masking strategies during pre-training, the mask span is set to 10 frames, and 8\% of the input representation is randomly selected as mask starting points for all models.

To pre-train the model with reconstruction-based targets and iterative clustering, we follow the HuBERT~\cite{hubert} training recipe. We use a batch size equivalent to 47 minutes of audio and train for 400k steps. The learning rate is linearly ramped up to 5e-4 over the first 32k steps, after which it decays linearly to zero over the remaining steps. For k-means clustering in iterative training, we implement an efficient clustering pipeline using \texttt{faiss}~\cite{douze2024faiss} toolkit, reducing clustering time from over a day to just a few hours. The k-means model is trained on a randomly sampled 10-hour subset of the full 960-hour dataset, using 500 clusters. We perform two rounds of iterative training.

For online clustering, we follow the setup of DinoSR~\cite{liu2024dinosr}. We use a batch size equivalent to 63 minutes of audio and train for 400k steps. Notably, masking is applied only to the input of the student model, while the teacher model processes unmasked input. 
Online clustering is performed on representations from layers 5 through 12 of the teacher model, where each layer is associated with a codebook of size 256. The codebook decay rate is fixed at 0.9. The learning rate schedule for the student model consists of a linear warm-up over the first 12k steps, followed by a constant phase for 188k steps, and a linear decay to zero over the remaining 200k steps.
The teacher model's exponential update rate starts at 0.999, increases to 0.9999 over the first 30k steps, and remains fixed for the subsequent 170k steps. After 200k steps, the teacher model is frozen, with the decay rate set to 1.0. Once pretraining is complete, we evaluate the student model on downstream tasks.

\subsection{Downstream Evaluation}
To assess the representations learned by Codec2Vec, we evalute its performance on the SUPERB benchmark~\cite{superb}. This benchmark includes a diverse set of speech tasks, including Phoneme Recognition (PR), Automatic Speech Recognition (ASR), Keyword Spotting (KS), Intent Classification (IC), Slot Filling (SF), Speaker Diarization (SD), Speaker Verification (SV), and Emotion Recognition (ER). 
We emphasize that the goal of our work is not necessarily to achieve higher accuracy on downstream tasks than these leading continuous-input models. Instead, we aim to demonstrate that Codec2Vec can practically match their performance while operating exclusively on highly compressed discrete units, which offers significant advantages in storage and training efficiency (as detailed in Section IV.B).
Therefore, we compare Codec2Vec against strong continuous-input baselines: DeCoAR 2.0~\cite{ling2020decoar}, HuBERT~\cite{hubert}, and DinoSR~\cite{liu2024dinosr}.
Additionally, to isolate the impact of substituting continuous waveforms with audio codec units inputs, we pretrain a model using the exact same training targets originally for pretraining HuBERT (denoted by \textsc{HuBERT-iter0-layer9-kmeans500})\footnote{https://github.com/espnet/espnet/tree/master/egs2/librispeech/ssl1} as a reference.

The results are presented in Table 1. 
First, comparing the standard waveform-input HuBERT with our Codec2Vec model (with~\textdagger), which is trained on the exact same target but uses discrete inputs, we observe that the our discrete-input model achieves highly competitive performance across all tasks.
This is a key finding, demonstrating that even with the inherent information loss from quantization, discrete audio codec units can effectively replace continuous signals as input for a strong SSL model like HuBERT without substantial performance degradation.
These results underscore the feasibility of using discrete codec units for general-purpose speech processing applications.

Next, we analyze the Codec2Vec models, which are built exclusively on discrete codec units for both input and varying target derivation strategies.
These models further highlight the capabilities of the Codec2Vec framework to learn effective representations directly from compressed data without relying on externally derived targets from continuous-input models.
The reconstruction-based Codec2Vec, which uses the input audio codec units themselves as prediction targets, establishes a useful performance baseline for this discrete-input framework.
Although it exhibits lower performance compared to continuous-input baselines like DeCoAR 2.0, it is worth noting that DeCoAR 2.0's methodology includes a vector quantization (VQ) step before spectrogram reconstruction, potentially encouraging more abstract learning targets.
Nonetheless, the performance gap between the reconstruction-based Codec2Vec and both DeCoAR 2.0 and cluster-based Codec2Vec variants (discussed below) suggests that codec units—while optimized for reconstruction fidelity—may be suboptimal as direct targets for general-purpose representation learning.

Consequently, when employing more sophisticated target derivation strategies, we observe that performance improves significantly.
The iterative clustering Codec2Vec, progressively refines training targets for subsequent round using k-means clustering on its own latent representations, achieves results that are competitive with the full waveform-based HuBERT on most tasks. Specifically, it outperforms HuBERT on SF, SD, and ER, while showing some degradation on ASR and IC.
Finally, the online clustering Codec2Vec also demonstrates strong performance, generally surpassing the waveform-based HuBERT and closely approaching the continuous-input DinoSR baseline, albeit with a slight gap.

Overall, these results confirm the feasibility of building self-supervised speech models that operate entirely on discrete audio codec units and achieve performance that is highly competitive with, and in several cases matches or even exceeds, strong continuous-input baselines.
The Codec2Vec framework, particularly with clustering-based targets, showcases the potential of discrete representations as a compelling and efficient alternative to conventional continuous acoustic inputs for general speech processing. 
While some performance trade-offs exist for specific tasks (e.g., ASR), the ability to practically match state-of-the-art continuous-input models using only a fraction of the data storage and pre-training time (detailed in Section IV.B ) is highly encouraging and represents a significant step towards more scalable and resource-efficient speech foundation models.


\subsection{Storage and Training Efficiency}

One of the key advantages of learning speech representations from discrete inputs, as facilitated by the Codec2Vec framework, is the significant reduction in both data storage requirements and the computational cost of SSL pre-training.
Table II shows a direct comparison using a HuBERT model trained for 400k steps on the 960-hour LibriSpeech dataset with identical targets (HuBERT-iter0-layer9-kmeans500), differing only in the input format (continuous waveform vs. pre-extracted discrete audio codec units).
Replacing continuous waveforms (stored as \texttt{.wav} files) with these pre-extracted units (stored as \texttt{.npz} files) reduces the dataset storage requirement from 60.4 GB to just 3.6 GB, a 16.5x reduction. This dramatic decrease not only conserves disk space but also enhances data transmission efficiency, particularly in low-bitrate or distributed training scenarios.
Furthermore, it enables entire large-scale speech datasets, which would typically require extensive disk I/O, to be loaded into RAM, substantially mitigating dataloading bottlenecks during training. For instance, the extensive LibriLight dataset (3.4 TB of raw audio) could potentially be compressed to approximately 12.6 GB of codec units in an ideal case (if using a designated data format for these discrete units), making in-RAM training feasible.

In terms of computational efficiency for the SSL pre-training phase, our discrete-input framework accelerates training by 2.3x (from 830 to 356 GPU hours). 
This speedup is measured in wall-clock GPU hours using the same hardware and experimental configuration for both continuous and discrete input models\footnote{While acknowledging theoretical FLOPs provide one view of computation, this wall-clock speedup reflects a more practical measure of real-world training efficiency gains}.
The reported training time of discrete input one includes the one-time cost for offline extraction of codec units from the LibriSpeech dataset, which amounted to approximately 6 GPU hours. 
The observed acceleration in pre-training primarily stems from two factors: (1) the elimination of the computationally intensive convolutional feature extractor typically required for processing raw waveforms or spectrograms, and (2) a significant reduction in I/O overhead due to the smaller data footprint, allowing more data to be cached or held in RAM.
Overall, These efficiency gains underscore Codec2Vec's potential as a practical and scalable solution, especially for resource-constrained environments or when pre-training on even larger unlabeled datasets where storage and compute time are critical bottlenecks.
\begin{table}[]
\caption{Comparison of storage requirements and computational costs for continuous vs. discrete inputs, using a HuBERT model trained on the 960-hour Librispeech dataset with \textsc{HuBERT-iter0-layer9-kmeans500} targets for 400k steps.}
\centering
\begin{tabular}{lcc}
\toprule
Input Type & Dataset Size & GPU Hours \\
\midrule
continuous & 60.4 GB   & 830                    \\
discrete (ours)   & \ \ 3.6 GB    & 356                    \\
\midrule
\textit{reduction ratio}      & \ \textit{16.5x}      & \ \textit{2.3x}    \\
\bottomrule \bottomrule
\end{tabular}
\end{table}

\begin{table}[]
\caption{Downstream performance comparison of using units from either DAC and Encodec as input for training a HuBERT model with \textsc{HuBERT-iter0-layer9-kmeans500}.}
\centering
\begin{tabular}{lccc}
\toprule
Input & PR & SF & SV \\
\midrule
DAC   & 5.2    & 88.6/24.5  & 5.1                  \\
EnCodec & 6.0 & 87.9/26.3  & 6.2\\
\bottomrule \bottomrule
\end{tabular}
\end{table}

\subsection{Impact of Neural Codec Model}
The choice of neural audio codec used to generate the discrete input units can significantly influence the performance of the downstream SSL model. 
To investigate this impact within the Codec2Vec paradigm, we pretrained a HuBERT model with discrete units derived from two different state-of-the-art neural audio codecs: DAC (the primary codec used in our main experiments) and an unofficial 16kHz variant of Encodec\footnote{Non-official 16kHz-variant: https://huggingface.co/pyp1/VoiceCraft}.
As shown in Table III, models trained with discrete units from DAC consistently outperform those using units from the Encodec variant across the evaluated tasks.
This performance gap highlights the critical influence of codec choice on representation learning, underscoring that the intrinsic characteristics of discrete units—shaped by the architecture, training objectives, and quantization strategies of the codec—substantially affect the quality of the learned SSL representations.
Notably, this also suggests that the selection of an appropriate existing codec could further enhance the performance of the Codec2Vec frameworks . 
Future work could explore optimizing neural audio codecs, particularly by refining their design to improve information preservation across diverse speech tasks while maintaining efficient compression.

\section{Discussion and Limitations}
Although our results confirm the feasibility and advantages of building a self-supervised speech model entirely on discrete inputs, several limitations remain: 
\begin{itemize} 
\item \textit{Codec Selection}: Neural audio codecs differ significantly in their ability to preserve various acoustic details influenced by factors such as training methodology, data domain, and quantization strategies.
As shown in prior works~\cite{codecSUPERB, mousavi2024dasb} and our experiments (Section III.C), this diversity directly impacts their ability to preserve acoustic details crucial for different downstream tasks.
Identifying or developing the optimal codec—whose discrete units are ideally suited for general-purpose speech representation learning—remains an open research question. 
\item \textit{Performance on Specific Tasks}: While Codec2Vec achieves competitive results across most evaluated tasks, certain tasks (e.g., ASR) consistently show performance trade-offs. 
These discrepancies might stem from the inherent information bottleneck of discrete units or the narrowband characteristics of codec models. Furthermore, it is important to recognize that current mainstream SSL pre-training strategies were predominantly developed for continuous input signals. Addressing these limitations may require not only further analysis of codec behavior but also the enhancement of existing SSL objectives and architectures.
\item \textit{Robustness to Noisy Conditions}: Although discrete codes inherently provide some robustness to noise, we have not extensively evaluated their performance in real-world noisy conditions.
The interaction between different noise types, codec compression, and the SSL pre-training process is complex.
A deeper investigation into the effects of various noise conditions on neural audio codecs is crucial to mitigating potential error propagation~\cite{tseng25_interspeech}.
\end{itemize}

\section{Conclusions}
In this work, we introduced \textbf{Codec2Vec}, the first self-supervised speech representation learning framework that relies exclusively on discrete units from neural audio codecs as input throughout pre-training.
Through a comprehensive evaluation of various masked prediction target derivation strategies, we demonstrated that Codec2Vec achieves performance on the SUPERB benchmark that is competitive with strong continuous-input baselines across a diverse range of speech processing tasks.
Critically, beyond this task performance, the Codec2Vec paradigm delivers substantial computational efficiencies, reducing data storage requirements by up to 16.5x and accelerating the SSL pre-training phase by up to 2.3x, establishing it as a compelling solution for large-scale pre-training and resource-constrained applications.
While some task-specific performance trade-offs remain, our findings robustly underscore the significant potential of discrete audio codec units as an efficient and effective alternative to traditional continuous representations for pre-training versatile speech foundation models, paving the way for more scalable and resource-aware speech processing.
Future work can further advance this paradigm by focusing on developing SSL objectives specifically tailored for discrete input sequences and by exploring neural audio codecs better-suited for learning effective representations for downstream tasks.


\section{Acknowledgements}
This material is based upon work supported by the National Science Foundation under Grant Number 2238605. Any opinions, findings, and conclusions or recommendations expressed in this material are those of the author(s) and do not necessarily reflect the views of the National Science Foundation.

\newpage
\bibliographystyle{IEEEtran}
\bibliography{mybib}


\end{document}